\documentclass[
aps, showpacs,preprintnumbers,amssymb,showkeys]{revtex4}
\usepackage{amssymb}
\usepackage{slashed}
\usepackage{amsmath}

\usepackage{graphicx}
\usepackage{dcolumn}
\usepackage{bm}

\def\nn{\noindent}
\newcommand{\be}{\begin{equation}}
\newcommand{\ee}{\end{equation}}
\newcommand{\bea}{\begin{eqnarray}}
\newcommand{\eea}{\end{eqnarray}}
\def\Re{{\cal R \mskip-4mu \lower.1ex \hbox{\it e}\,}}
\def\Im{{\cal I \mskip-5mu \lower.1ex \hbox{\it m}\,}}
\def\ie{{\it i.e.}}

\def\ibid{{\it ibid}.}
\def\tev{\,{\ifmmode\mathrm {TeV}\else TeV\fi}}
\def\gev{\,{\ifmmode\mathrm {GeV}\else GeV\fi}}
\def\mev{\,{\ifmmode\mathrm {MeV}\else MeV\fi}}
\def\to{\rightarrow}

\begin{document}
\title{Higgsstrahlung and pair production in $e^{+}e^{-}$ collision in the noncommutative standard model}
\author{Weijian Wang, Feichao Tian, and Zheng-Mao Sheng\footnote{Corresponding author's email:zmsheng@zju.edu.cn}}
\affiliation{Department of Physics, Zhejiang University, Hangzhou
310027, China}

\begin{abstract}
 The Higgsstrahlung process $e^+e^-\to Z H$ and pair production process
$e^+e^- \to H H$ are studied in the framework of the minimal
noncommutative (NC) standard model. In particular, the Feynman
rules involving all orders of the noncommutative parameter
$\theta$ are derived using reclusive formation of Seiberg-Witten
map. It is shown that the total cross section and angular
distribution can be significantly affected because of spacetime
noncommutativity when the collision energy exceeds to 1 \tev. It
is found that in each process, there is an optimal collision
energy ($E_{oc}$) for achieving the greatest noncommutative
effect, and $E_{oc}$ varies linearly with the NC scale
$\Lambda_{NC}$. A brief discussion on the process $e^+e^- \to\mu^+
\mu^-$ is also given.

\end{abstract}

\maketitle

\section{INTRODUCTION}
In string theory, noncommutative (NC) space-time appears naturally
in D-brane dynamics in the low energy limit
\cite{Connes98,Douglas98,Seiberg99}. It is generally believed that
the stringy effect can only be observed at the Plank scale
$M_{P}$, which is at far from detectable. However, given the
possibility \cite{c11,Hewett01} that the large hierarchy between
the gravitational scale $M_{P}$ and the weak scale $M_{W}$ can be
narrowed down to a few \tev, one can expect to see the NC effect
predicted by the noncommutative field theory (NCQFT)at around 1
\tev. The noncommutative space-time can be characterized by the
coordinate
 operators satisfying

\begin{equation}
[\hat{x}_{\mu},\hat{x}_{\nu}]=i\theta_{\mu\nu}=\frac{ic_{\mu\nu}}{\Lambda^2_{NC}},
\label{homer}
\end{equation}
where the matrix $\theta_{\mu\nu}$ in Eq.~(\ref{homer}) is
constant, antisymmetric and real. The elements of the
dimensionless constant matrix $c_{\mu\nu}$ are assumed to be of
order unity and $\Lambda_{NC}$ represents the NC scale, having the
dimension of inverse mass. NCQFT can be constructed though Weyl
correspondence, where the ordinary product of fields is replaced
by the Moyal-Weyl star product \cite{Seiberg99}
\begin{equation}
(f\star
g)(x)=\exp(\frac{1}{2}\theta_{\mu\nu}\partial_{x^{\mu}}\partial_{y^{\nu}})f(x)f(y)|_{y=x}.
\label{marge}
\end{equation}
Using this method,  high energy processes of quantum
electrodynamics in noncommutative space-time (NCQED)
 have been extensively studied \cite{Hewett01,jft01}. An interesting consequence is
 the raising of triple and 4-point photon vertex in NCQED analogous to the Yang-Mills gauge theory. However, some obstructions
 such as charge quantization\cite{Hayakawa00} and no-go theorem\cite{Chai02} must be considered
if one in tends to build an arbitrary gauge theory. Up to now
there are two versions of the noncommutative standard model
(NCSM). One is that the gauge group is restricted to
$U(3)*U(2)*U(1)$ \cite{Chai03}. In this case, however, additional
heavy gauge bosons and a delicate Higgs mechanism has to be
introduced in order to remove two extra $U(1)$ factors. Another is
a minimal version of the noncommutative standard model
(mNCSM)\cite{Calmet07}, in which the group closure property is
still valid when one generalizes the SU(3)*SU(2)*U(1) Lie algebra
gauge theory to the enveloping algebra value using the
Seiberg-Witten map (SWM) method\cite{Seiberg99}.

The SWM means that both the matter fields $\hat{\psi}$ and the
gauge fields $\hat{A}_\mu$ in noncommutative space-time can be
expanded in terms of the commutative ones as power series in
$\theta$,
\begin{eqnarray}
\hat{\psi}(x,\theta)&=&\psi(x)+\theta\psi^{(1)}+\theta^2\psi^{(2)}+...  \label{bart}                      \\
\hat{A}_{\mu}(x,\theta)&=&A_{\mu}(x)+\theta
A^{(1)}_{\mu}+\theta^2A^{(2)}_{\mu}+... \label{lisa}
\end{eqnarray}

The striking feature of mNCSM is that it predicts new physics
which are not only the noncommutative correction of particle
vertices but also new interactions
 beyond the SM in ordinary space-time. This attracts many authors to
focus attention on the noncommutative phenomenology of particles
based on mNCSM. Recently, several high energy processes such as
$e^{-} e^{-}\to e^+ e^-$ (Moller), $e^+ e^- \to e^+ e^-$ (Babaha)
\cite{Das08},
 $e^{+}e^{-} \to \gamma\gamma$ \cite{Alboteanu07}, $e^{+}e^{-}\to \mu^+\mu^-$ \cite{Ab10} and neutrino-photon
 scattering \cite{Haghighat06} have been investigated in the context
 of mNCSM. The possibility to detect NC effect though SM forbidden decay such as $Z \to \gamma\gamma$ \cite{Buric07},
 $J/\psi \to \gamma\gamma$ and $ K\to \pi\gamma$ \cite{Melic05} have also
been explored by many authors in order to obtain a lower
$\Lambda_{NC}$  constraint.

Most of the existing analysis are only up to the first $\theta$
order. It is necessary to examine higher order contributions since
in future colliders the center mass energy can be comparable or
even exceed the NC scale. In a recent work \cite{Horvat11}, the
authors pointed out that an incorrect $\Lambda_{NC}$ lower bound
could be obtained from ultra-high energy cosmic ray experiments if
one simply expands the noncommutative interaction term to the
linear order. To overcome this, the $\theta$-exact expression of
SWM was derived by directly solving the gauge equivalence relation
and applied to ultra-high energy neutrino processes
\cite{Horvat112}.

On the other hand, the Higgs boson, although not yet observed can
play an important role in electroweak spontaneous symmetry
breaking (SSB) through which the gauge boson can have mass. The
LEP2 experiment gives a lower bound of 114.4 \gev \cite{Al02} and
if we take the Global Electro-weak fit into account,
 the Higgs mass should be no more than 200 \gev \cite{Djouadi96}.
Recently, The CDF and D0 collaborations at the Tevatron exclude
the Higgs boson in the range between 158 \gev and 175 \gev at 95
confidence level \cite{CDF}. It is believed that the collider such
as LHC and the planned International Linear Collider(ILC) will
help
 people to prove or exclude the existence of Higgs boson.

It is interesting to see if new physics can appear along with the
Higgs boson. The possibility has already been  extensively
discussed in many theories beyond the SM. In this paper, we
explore the higgsstrahlung process: $e^+e^-\to H Z$, and the pair
production process: $e^+e^- \to H H$ in the framework of mNCSM.
The later channel is forbidden in ordinary SM and has been studied
recently\cite{Das11} in the linear $\theta$ order. However, the
results of Ref. \cite{Das11} are not valid when the on-shell
condition is applied. In Sec. II, the n-th order SWM solution is
given as a recursive formulation from  the Seiberg-Witten
differential equation. Although the resulting expression is
lengthy, most terms are not relevant to the interaction
considered, thus allowing us to derive the full-$\theta$
expression for the fermion, gauge boson, and Higgs boson. In Sec.
III, We give the scattering amplitudes of $e^+e^-\to H Z$ and
$e^+e^-\to H H$. We shall also briefly discuss the process
$e^+e^-\to\mu^+ \mu^-$. Numerical analysis of total cross section
and azimuthal angular distribution of cross section are presented
in Sec. IV. Finally, we summarize and discuss our results in Sec.
V.

\section{SEIBERG-WITTEN MAPS AND NONCOMMUTATIVE STANDARD MODEL}
SWM relates the noncommutative fields to their counterpart in
ordinary space-time. When the limit $\theta \to0$  is taken, the
noncommutative fields reduce to the ordinary ones in commutative
space-time. SWM can be derived as perturbative solutions of the
gauge equivalence relation order by order. It is shown in Ref.
\cite{Ulker08} that the n-th order SWM can also be obtained from a
differential equation introduced by Seiberg and
Witten\cite{Seiberg99}. The SW-differential equation of the gauge
field $\hat{V}_{\mu}$ is \cite{Seiberg99,Ab10}

\begin{equation}
\delta\theta^{\kappa\lambda}\frac{\partial\hat{V}_\mu}{\partial\theta^{\kappa\lambda}}
=-\frac{1}{4}\delta\theta^{\kappa\lambda}\{\hat{V}_{\kappa},
\partial_{\lambda}\hat{V}_{\mu}+\hat{F}_{\lambda\mu}\}_{*},\label{maggie}
\end{equation}
and that of the fermion fields $\hat{\Psi}$ is \cite{Ulker08}

\begin{equation}
\delta\theta^{\kappa\lambda}\frac{\partial\hat{\Psi}}{\partial\theta^{\kappa\lambda}}
=-\frac{1}{4}\delta\theta^{\kappa\lambda}\hat{V}_{\kappa}*(\partial_{\lambda}\hat{\Psi}+\hat{D}_{\lambda}\hat{\Psi}),
\label{akira}
\end{equation}
which can be derived  by changing $\theta $ to $ \theta +
\delta\theta$ . After inserting Eqs. \eqref{maggie} and
\eqref{akira} to the Taylor expansions of the NC fields, the n-th
order solution can be obtained. Here we  list the  results given
in Refs. \cite{Ulker08,Bichl01},

\begin{eqnarray}
\hat{V}^{(n+1)}_{\mu}&=&-\frac{1}{4(n+1)}\theta^{\kappa\lambda}
\sum_{\alpha+\beta+\gamma=n}\{\hat{V}^{(\alpha)}_\kappa,
\partial_\lambda \hat{V}^{(\beta)}_{\gamma}+\hat{F}^{(\beta)}_{\lambda\gamma}\}_{*^{(\gamma)}}. \label{albright}\\
\hat{\Psi}^{(n+1)}&=&-\frac{1}{4(n+1)}\theta^{\kappa\lambda}\sum_{\alpha+\beta+\gamma=n}\hat{V}^{(\alpha)}_\kappa
*^{(\gamma)} (\partial_\lambda
\hat{\Psi}^{(\beta)}+(D_{\lambda}\hat{\Psi})^{(\beta)}).
\label{aristotle}
\end{eqnarray}
Following Ref. \cite{Calmet07}, the fermion, and Higgs and Yukawa
sectors of mNCSM are

\begin{eqnarray}
S_{fermions}&=&\int d^4x\sum_{i=1}^3(\bar{\hat{l}}_{L}^{(i)}*(i\hat{\slashed{D}}\hat{l}_L^{(i)})
+ \bar{\hat{Q}}_L^{(i)}*(i\hat{\slashed{D}}\hat{Q}_L^{(i)})  \nonumber \\
& &+\bar{\hat{l}}_R^{(i)}*(i\hat{\slashed{D}}\hat{l}_R^{(i)}) +
\bar{\hat{u}}_R^{(i)}*(i\hat{\slashed{D}}\hat{u}_R^{(i)}) +
\bar{\hat{d}}_R^{i}*(i\hat{\slashed{D}}\hat{d}_R^{(i)}))
\label{fernando}
\end{eqnarray}

\begin{equation}
S_{Higgs}=\int d^4x
[(\hat{D}_{\mu}\hat{\Phi})^{\dagger}*(\hat{D}^\mu\hat{\Phi})-\mu^2\hat{\Phi}^{\dagger}*
\hat{\Phi}-\lambda\hat{\Phi}^\dagger*\hat{\Phi}*\hat{\Phi}^\dagger*\Phi]
\label{atkins}
\end{equation}

\begin{eqnarray}
S_{Yukawa}&=&-\int d^4x \sum_{i,j=1}^3 [C_l^{(ij)}(\bar{\hat{l}}_L^{(i)}*\Phi_l*\hat{e}_R^{(j)})
 + C_l^{\dagger(ij)}(\bar{\hat{e}}_R^{(i)}*{\hat{\Phi}}_L^\dagger*\hat{l}_L^{(j)})  \nonumber \\
 & &+C_{u}^{(ij)}(\bar{\hat{Q}}_L^{(i)}*\hat{\Phi}_u^c*{u}_R^{(j)})
 + C_u^{\dagger(ij)}(\bar{\hat{u}}_R^{(i)}*\hat{\Phi}_u^{c^\dagger}*\hat{Q}_L^{(j)})  \nonumber \\
  & &+C_d^{(ij)}(\bar{\hat{Q}}_L^{(i)}*\hat{\Phi}_d*\hat{d}_R^{(j)})
  + C_d^{\dagger(ij)}(\bar{\hat{d}}_R^{(i)}*\Phi_d^\dagger*\hat{Q}_L^{(j)}) ]  \label{birch}
\end{eqnarray}
with
\begin{equation}
Q=\binom{u}{d}, \quad\quad\quad\quad
l=\binom{\nu}{e},\quad\quad\quad\quad \Phi^c=i\tau_2\Phi^*,
\end{equation}
where $\nu$, $e$, $u$, $d$, $l$ and $Q$ stand for the neutrinos,
charged leptons, up-type quarks, down-type quarks, lepton doublets
and quark doublets, respectively, for three generations (To avoid
confusion, we denote electron by $e^{-}$),  and the subscripts $L$
and $R$ stand for the left- and right-hand, respectively. The
expression given above is the same as the SM in ordinary
space-time, except for the replacement of the ordinary fields by
corresponding NC fields and substitution of the ordinary product
by the star products\cite{Calmet07}. Note that $\hat{\Phi} $ and
$\hat{\Phi}_Y$ $(Y=l,u,d)$ are the noncommutative Higgs fields in
the free  and Yukawa sectors, respectively. The NC Higgs field
$\hat{\Phi}_Y$ transforms under two different gauge groups. The
corresponding gauge potentials $\hat{V}_\mu $ and $\hat{V}'_\mu$
inherited from the fermions on the right and left of the Higgs
fields in Yukawa sector. Thus, the SWM of $\hat{\Phi}_Y$ has a
hybrid feature and is given by

\begin{eqnarray}
\hat{\Phi}_Y &\equiv& \hat{\Phi}[\Phi. V,V'] \label{Paul} \\
\nonumber &=& \Phi +
\frac{1}{2}\theta^{\kappa\lambda}V_{\lambda}(\partial_{\kappa}\Phi
- \frac{i}{2}(V_{\kappa}\Phi-\Phi V'_\kappa))\\ \nonumber & &
+\frac{1}{2}\theta^{\kappa\lambda}(\partial_{\kappa}\Phi-
\frac{i}{2}(V_{\kappa}\Phi - \Phi
V'_{\kappa}))V'_{\lambda}+\mathcal{O}(\theta^2)\nonumber
\end{eqnarray}

The hybrid SWM guarantees the equivalence of covariant
transformation between the noncommutative and ordinary fields,
which means
\begin{equation}
\delta_{\lambda,\lambda'}\hat{\Phi}_Y[\Phi_Y,V,V']
=i\hat{\Lambda}*\hat{\Phi}_Y-i\hat{\Phi}_Y*\hat{\Lambda}',\label{Robert}
\end{equation}
where $\hat{\Lambda}$, $\hat{\Lambda}'$ are noncommutative gauge
parameters corresponding to their ordinary counterparts ($\lambda$
and $\lambda'$). In Ref. \cite{Calmet07}, the representation of
SMW for NC Higgs field $\hat{\Phi}$ in the Higgs kinetic sector
Eq. $\eqref{atkins}$ is $\hat{\Phi}\equiv
\hat{\Phi}[\Phi,V_\mu,0]$, which is chosen to be of the same
representation as the standard model. From the point of gauge
invariance, however, there is no a priori requirement that we must
take this simplest representation. In order to explore as much new
physics as possible, here we choose a more general SWM expression

\begin{equation}
\hat{\Phi}\equiv \hat{\Phi}[\Phi , V_\mu, V'_\mu]\label{Robin}
\end{equation}
where
\begin{eqnarray}
V_\mu & = & x g' B_\mu + gW^a_\mu\frac{\sigma^a}{2}, \label{Stacy}\\
V_\mu'& = & -(\frac{1}{2}-x)g'B_\mu, \label{Marshall}
\end{eqnarray}
and $B_{\mu}$, $W_{\mu}^a$ and $g'$, $g$ are gauge fields and
coupling constants of the $U(1)$ and $SU(2)$ groups, respectively,
in the usual space-time. The parameter $x$ introduced here
represents the ambiguity of noncommutative U(1) gauge transform
which means that the covariant derivative for $\hat{\Phi}$ thus is
given by

\begin{equation}
\hat{D}_\mu\hat{\Phi}=\partial_\mu\hat{\Phi}-i\hat{V}_\mu*\hat{\Phi}
+ i\hat{\Phi}*\hat{V}_\mu'\label{Peter}
\end{equation}
Clearly, this formulation reduces to the commutative one with
right hypercharge in SM if one sets $\theta \to 0$. Now we derive
the n-th order SWM for the Higgs field. Following Ref.
\cite{Ulker08,Bichl01}, we get the SW-differential equation

\begin{equation}
\delta\theta^{\kappa\beta}\frac{\delta\hat{\Phi}}{\delta\theta^{\kappa\lambda}}
=-\frac{1}{2}\delta\theta^{\kappa\lambda}[\hat{V}_{\kappa}*(\partial_{\lambda}\hat{\Phi}
- \frac{i}{2}(\hat{V}_{\lambda}*\hat{\Phi} -
\hat{\Phi}*\hat{V'}_{\lambda})) -
\frac{1}{2}(\partial_{\kappa}\hat{\Phi} -
\frac{i}{2}(\hat{V}_{\kappa}*\hat{\Phi} -
\hat{\Phi}*\hat{V'}_{\kappa}))*\hat{V'}_{\lambda}]\label{Phillip}
\end{equation}
which can be written as
\begin{equation}
\begin{split}
\frac{\delta\hat{\Phi}}{\delta\theta^{\kappa\lambda}}&=-\frac{1}{4}\hat{V}_{\kappa}*
(\partial_{\lambda}\hat{\Phi}-\frac{i}{2}(\hat{V'}_{\lambda}*\hat{\Phi}
 - \hat{\Phi}*V'_{\lambda})) + \frac{1}{4}\hat{V}_{\lambda}*(\partial_{\kappa}\hat{\Phi}
  - \frac{i}{2}(\hat{V}_{\kappa}*\hat{\Phi} - \hat{\Phi}*\hat{V'}_{\kappa}))\\
& -
\frac{1}{4}(\partial_{\lambda}\hat{\Phi}-\frac{i}{2}(\hat{V}_{\lambda}*
\hat{\Phi}-\hat{\Phi}*\hat{V'}_{\lambda}))*\hat{V'}_{\kappa}
+ \frac{1}{4}(\partial_{\kappa}\hat{\Phi} -
\frac{i}{2}(\hat{V}_{\kappa}\hat{\Phi} -
\hat{\Phi}*\hat{V'}_{\kappa}))*\hat{V'}_{\lambda}  \label{Muse}
\end{split}
\end{equation}

On the other hand,  $\hat{\Phi}$ can be  Taylor expanded to up to
the $(n+1)\quad\theta$th order in $\theta$
\begin{equation}
\begin{split}
\hat{\Phi}^{n+1}&=\hat{\Phi}^{(0)}+\hat{\Phi}^{(1)}+...+\hat{\Phi}^{(n+1)}\\
&=\Phi +
\sum_{k=1}^{n+1}\frac{1}{k!}\theta^{\mu_1\nu_1}...\theta_{\mu_\kappa\nu_\kappa}
(\frac{\partial^k}{\partial\theta^{\mu_1\nu_1}...
\partial\theta^{\mu_\kappa\nu_\kappa}}\hat{\Phi}^{(n+1)})_{\theta=0}
\label{Frank}
\end{split}
\end{equation}

Inserting Eq. \eqref{Muse} into Eq. \eqref{Frank}, one can get the
recursive solution up to the n+1 order
\begin{equation}
\begin{split}
\hat{\Phi}^{(n+1)}&=-\frac{1}{4(n+1)}\theta^{\kappa\lambda}\sum_{\alpha+\beta+\gamma=n}
[\hat{V}_{\kappa}^{(\alpha)}*^{(\beta)}(\partial_{\lambda}\hat{\Phi})^{(\gamma)}
 + \hat{V}_{\kappa}^{(\alpha)}*^{(\beta)}(\hat{D}_{\lambda}\hat{\Phi})^{(\gamma)} \\
&+
(\partial_{\lambda}\hat{\Phi})^{(\alpha)}*^{(\beta)}\hat{V}_{\kappa}^{(\gamma)}+
(\hat{D}_{\lambda}\hat{\Phi})^{(\alpha)}*^{(\beta)}{\hat{V}}'^{(\gamma)}_\kappa],
\label{Phin}\end{split}
\end{equation}
where
\begin{equation}
(\hat{D}_{\lambda}\hat{\Phi})^{(\gamma)}\equiv(\partial_{\lambda}\hat{\Phi})^{(\gamma)}
-i\sum_{m+n+t=\gamma}(\hat{V}^{(m)}_{\lambda}*^{(n)}\hat{\Phi}^{(t)}
-\hat{\Phi}^{(t)}*^{(n)}\hat{V}^{'(m)}_{\lambda})
\label{cai}
\end{equation}

\begin{figure}
\begin{minipage}[t]{0.5\linewidth}
\centering
\includegraphics[width=2.2in]{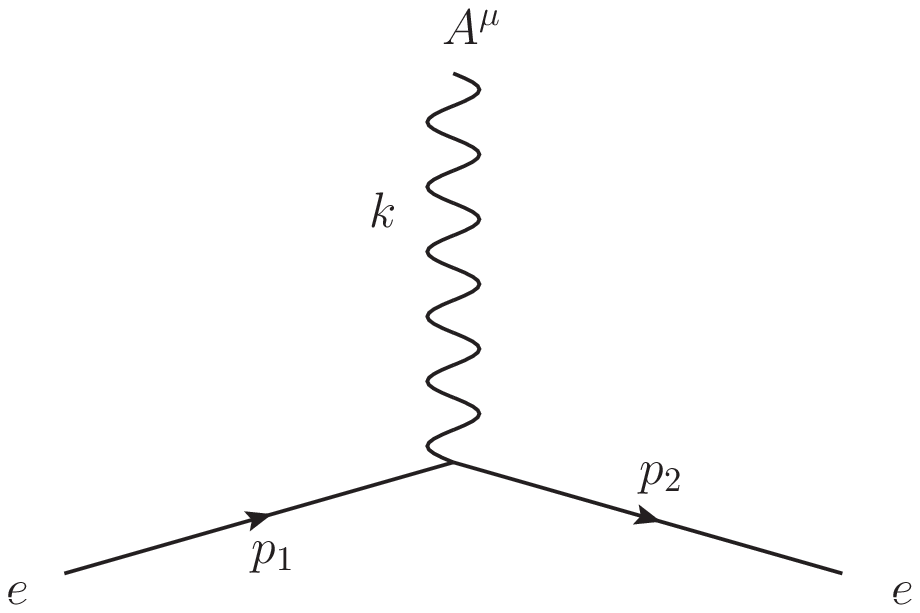}
\caption{$\gamma$-e-e\label{f1}}
\end{minipage}%
\begin{minipage}[t]{0.5\linewidth}
\centering
\includegraphics[width=2.2in]{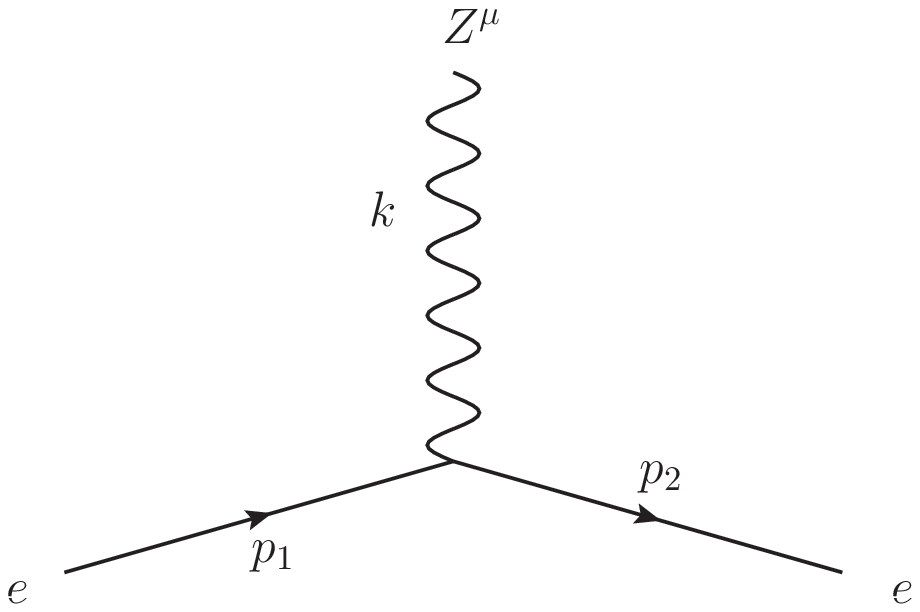}
\caption{Z-e-e\label{f2}}
\end{minipage}
\begin{minipage}[t]{0.5\linewidth}
\centering
\includegraphics[width=2.2in]{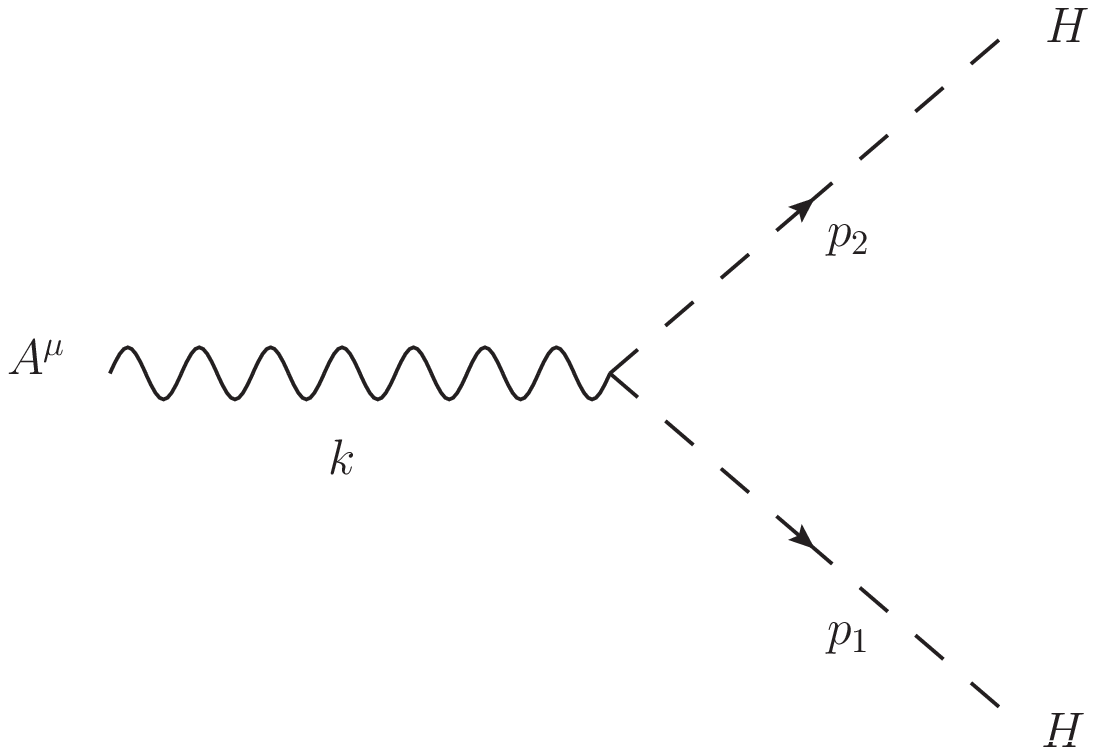}
\caption{$\gamma$-H-H\label{f3}}
\end{minipage}%
\begin{minipage}[t]{0.5\linewidth}
\centering
\includegraphics[width=2.2in]{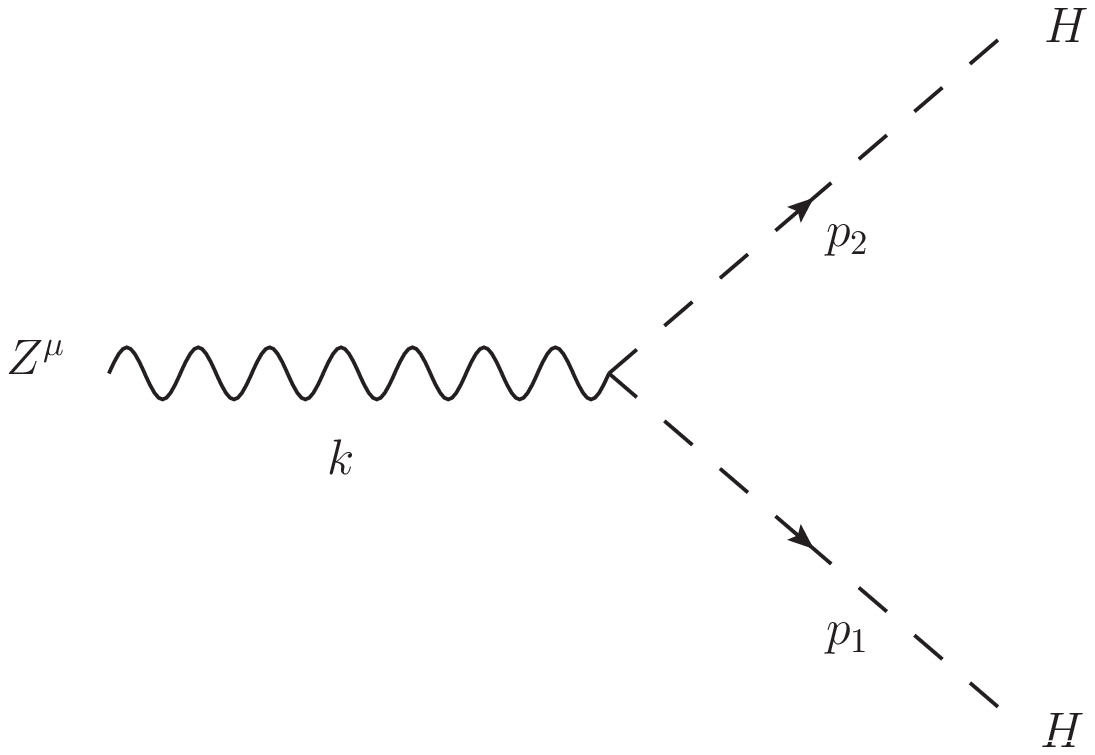}
\caption{Z-H-H\label{f4}}
\end{minipage}
\begin{minipage}[t]{0.5\linewidth}
\includegraphics[width=2.2in]{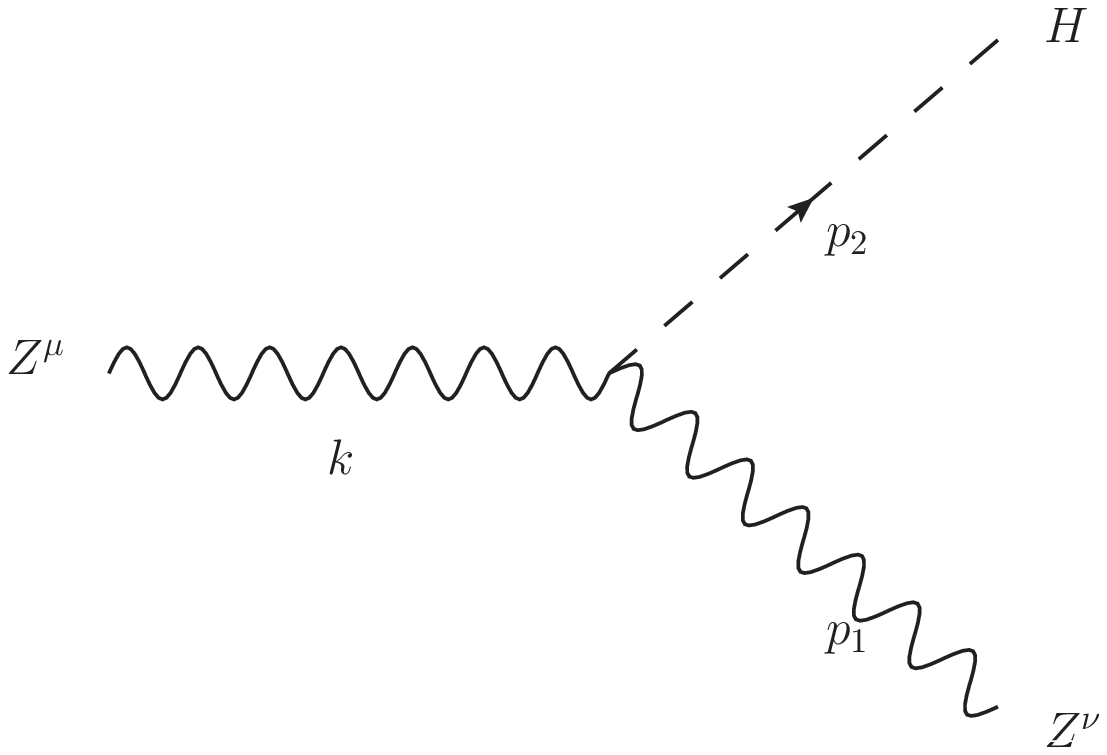}
\caption{Z-H-Z\label{f5}}
\end{minipage}%
\end{figure}

The all-expanded expressions of Eqs. \eqref{albright},
\eqref{aristotle} and \eqref{Phin} are rather lengthy, so that in
many  existing works they are limited to the lowest $\theta$
order. We note, however, that for the high energy process
discussed in this paper,
 the number of the gauge and matter fields taking part in each particle vertex
 is no more than two. The terms with three or more gauge
fields can thus be  set to zero and the solutions of SWM can then
be written in a compact form
\begin{eqnarray}
\hat{V}_{\mu}^{(\text{eff})}&=&V_{\mu} - \frac{1}{4}\theta^{\kappa\lambda}V_{\kappa}(\hat{O}
+ \hat{O}')(\partial_{\lambda}V_{\mu} + F_{\lambda\mu}), \label{Scofield}\\
\hat{\Psi}^{(\text{eff})}&=&\psi-\frac{1}{2}\theta^{\kappa\lambda}
V_{\kappa}\hat{O}(\partial_{\lambda}\psi),  \label{Burrows}\\
\hat{\Phi}^{(\text{eff})}&=&\Phi -
\frac{1}{2}\theta^{\kappa\lambda}V_{\kappa}\hat{O}(\partial_{\lambda}\Phi)
-\frac{1}{2}V'_{\kappa}\hat{O}'(\partial_{\lambda}\Phi) +
\frac{i}{4}\theta^{\kappa\lambda}(V_{\kappa}\hat{O}V_{\lambda}^{\Phi})\Phi
\label{Tan}\\ \nonumber & &+
\frac{i}{4}\theta^{\kappa\lambda}(V'_{\kappa}\hat{O}'V_{\lambda}^{\Phi})\Phi
+ \hat\Phi(V,V',\partial\Phi),
\end{eqnarray}

where
\begin{eqnarray}
\hat{O}&\equiv&\frac{e^{\frac{i}{2}\theta^{\alpha\beta}\overleftarrow{\partial}_{\alpha}
\overrightarrow{\partial}_{\beta}}-1}{\frac{i}{2}\theta^{\alpha\beta
\overleftarrow{\partial}_{\alpha}\overrightarrow{\partial}_{\beta}}}, \\
\hat{O}'&\equiv&\hat{O}|_{\theta\to-\theta},\\
V_{\mu}^{\Phi}&\equiv&V_{\mu}-V_{\mu}'.
\end{eqnarray}
The superscript "eff" means that we only keep the terms taking
part in the process $ e^+e^- \to H Z$ and $e^+e^-\to H H$. The
last term on the RH side of Eq. \eqref{Tan} contains two gauge
fields and derivatives of $\Phi$. Inserting Eqs. \eqref{Scofield},
\eqref{Burrows} and \eqref{Tan} into Eqs. \eqref{fernando} and
\eqref{atkins} and imposing  spontaneous symmetry breaking under
the unitary gauge
\begin{equation}
\Phi=\binom{\Phi^{\dagger}}{\Phi^{0}}\xrightarrow{SSB}\frac{1}{\sqrt{2}}\binom{0}{v+h}
\end{equation}
where $v$ is the vacuum expectation value, we derive the relevant
vertex and Feynman rules. We cannot give an non-perturbative
expression for the term $\hat{\Phi}(V,V',\partial{\Phi})$.
However, it is easy to verify that the contribution of this term
to the interaction under consideration is zero.
 We show the vertex needed for processes $ e^+e^- \to H Z$ and $e^+e^-\to H H$ in Figs. 1-5
 where all the gauge boson momenta are ingoing except that for  $p_{1}$
 in Fig. \ref{f5}. The relative Feynman rules are

\begin{equation}
V^{\mu}_{1}(p_1, k, p_2)= ie\gamma^{\mu}e^{\frac{i}{2}p_1\theta
p_2}
\end{equation}
for photon-charged lepton vertex,
\begin{equation}
V^{\mu}_{2}(p_1, k,
p_2)=-\frac{ie}{\sin2\theta_{W}}\gamma^{\mu}(C_V-C_A\gamma_5)e^{\frac{i}{2}p_1\theta
p_2}
\end{equation}
for Z boson-charged lepton vertex,
\begin{equation}
V^{\mu}_{3}(p_1, k,
p_2)=2e(x-\frac{1}{2})(p_2-p_1)^{\mu}\sin(\frac{1}{2}p_1\theta
p_2)
\end{equation}
for photon-Higgs-Higgs vertex,
\begin{equation}
V^{\mu}_{4}(p_1, k,
p_2)=2e[(x-\frac{1}{4})\tan\theta_{W}+\frac{1}{4}\cot\theta_{W}](p_2-p_1)^{\mu}\sin(\frac{1}{2}p_1\theta
p_2)
\end{equation}
for Z boson-Higgs-Higgs vertex, and
\begin{equation}
V^{\mu}_{5}(p_1, k,
p_2)=\frac{iem_Z}{\sin2\theta_W}[2\cos(\frac{1}{2}p_1\theta
p_2)g_{\mu\nu}+\frac{1}{4}((\theta p_2)_\mu p_{1\nu}+(\theta
p_2)_{\nu}k_{\mu}).(\frac{\cos(\frac{1}{2}p_1\theta
p_2)-1}{p_1\theta p_2})]
\end{equation}
for Z boson-Z boson-Higgs vertex. Here,
$C_V=-\frac{1}{2}+2\sin^2\theta_W$, $C_A=-\frac{1}{2}$, and
$\theta_{W}$ is the Weinberg angle. The masses of the Higgs, Z,
and W bosons can be written as
\begin{equation}
\begin{split}
m^2_H&= -2\mu^2 = 2v^2\lambda, \\
m^2_W&=\frac{1}{4}v^2g^2, \quad\quad\quad\quad\quad
m^2_Z=\frac{1}{4}v^2(g^2+{g'}^2)=\frac{m^2_W}{\cos^2\theta_W}.
\end{split}
\end{equation}
Since we are only concerned with the lowest tree level process, we
apply the equations of motion to the particles in external line,
and ignore the terms vanishing due to on-shell condition. It
should be mentioned that the Feynman rule for Z-H-H above is
different from the one given in Ref. \cite{Das11} even at the
$\theta$ order. The detailed calculation is in the Appendix. It is
found that the Feynman rules in Ref. \cite{Das11}are not complete,
and cannot work for the on-shell condition.

\section{SCATTERING AMPLITUDES IN NCSM}

\subsection{$e^+ e^-\to Z H$}
\begin{figure}
 \includegraphics[scale = 0.56]{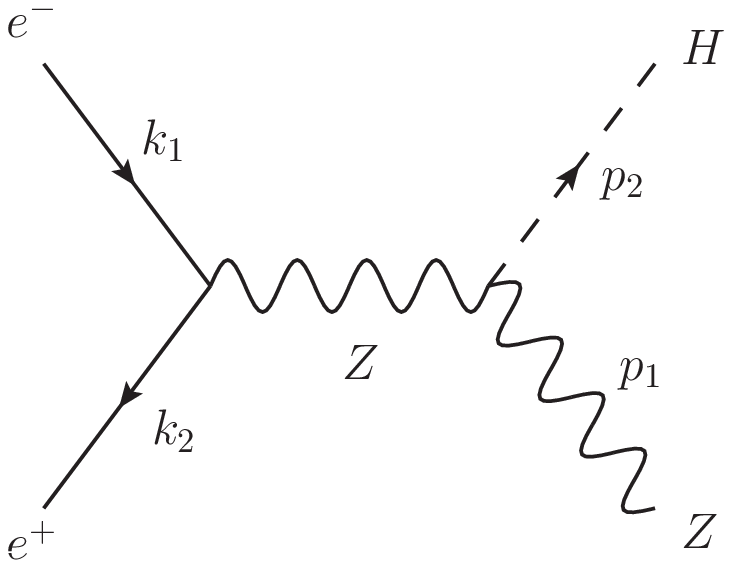}
 \caption{Feynman diagrams for process $e^+e^-\to Z H$\label{eq11fey}}
 \end{figure}

The tree level Feynman diagram is shown in Fig. \ref{eq11fey},
where the momenta $p_1$ of Z boson external line is outgoing. The
process is s-channel and proceeds though mediated $Z$ bosons.
Using the Feynman rules in Sec. II, the relative amplitude is
given by

\begin{equation}
M=\frac{i
e^2m_{Z}}{\sin^22\theta_{W}}\bar{v}(k_2)\gamma_{\mu}(C_{V}-C_{A}\gamma_5)u(k_1)\frac{i}{s-m_{Z}^2+i\Gamma_{Z}}\Gamma^{\mu\nu}(p_1,
p_2)\epsilon_{\nu}^*(p_1)e^{\frac{i}{2}k_2\theta k_1},
\end{equation}
in which
\begin{equation}
\Gamma^{\mu\nu}(p_1,p_2)=2\cos(\frac{1}{2}p_1\theta
p_2)g^{\mu\nu}+\frac{1}{4}[\frac{\cos(\frac{1}{2}p_1\theta
p_2)-1}{p_1\theta p_2}]\times[(\theta p_2)^{\mu}p_1^{\nu}+(\theta
p_2)^{\nu}k^{\mu}],
\end{equation}
were $k_1$, $k_2$, $p_1$ and $p_2$ are four momentums of electron,
positron, Higgs boson and outgoing Z boson; $s_1,s_2$ are spin
indices, $s=(k_1+k_2)^2=(p_1+p_2)^2$, and $\Gamma_Z$ is the decay
width of the $Z$ boson. We omit the electron (positron) mass in a
high energy limit.

\subsection{$e^+e^-\to H H$}

\begin{figure}
 \includegraphics[scale = 0.56]{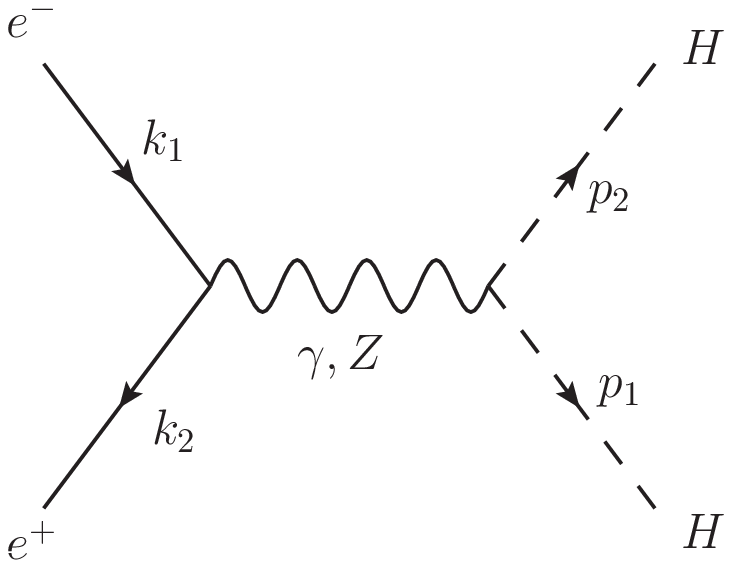}%
 \caption{Feynman diagrams for process $e^+e^-\to H H$\label{eq22fey}}
 \end{figure}

We now  give the scattering amplitude of neutral Higgs boson pair
production. The corresponding Feynman diagrams of the process are
shown in Fig. \ref{eq22fey}. Different from higgsstrahlung, this
process is forbidden in ordinary SM.
 Using the Feynman rules given in Sec. II, we give the following amplitudes,

\begin{equation}
M_{\gamma}=\frac{2e^2}{s}(x-\frac{1}{2})\bar{v}(k_2)\gamma^{\mu}u(k_1)(p_1-p_2)_{\mu}\sin(\frac{1}{2}p_1\theta
p_2)e^{\frac{i}{2}k_2\theta k_1}\label{tjack}
\end{equation}
for $\gamma$ mediated and
\begin{equation}
\begin{split}
M_{Z}&=-\frac{2e^2}{\sin 2\theta_{W}}[(x-\frac{1}{4})\tan\theta_u
+ \frac{1}{4}\cot\theta_{W})]\bar{v}(k_2)\gamma^{\mu}(C_V-C_A\gamma_5)u(k_1) \\
 &\times\frac{1}{s-m_{Z}^2+i\Gamma_{Z}m_{Z}}(p_1-p_2)_{\mu}\sin(\frac{1}{2}p_1\theta p_2)e^{\frac{i}{2}k_2\theta k_1}\label{john}
\end{split}
\end{equation}
for $Z$ mediated. The total amplitude is
\begin{equation}
M=M_\gamma + M_{Z}.\label{sayid}
\end{equation}

\subsection{$e^+e^-\to \mu^+\mu^-$}

 Using the SWM
 expanding to the $\theta^2$ order, the squared-amplitude for $e^+ e^- \to \mu^+ \mu^-$ up to the $\theta^4$
order was studied in Ref. \cite{Ab10}. An interesting result is
that all the contribution from $\theta, \theta^2$ and $\theta^3$
terms to the cross section cancelled out. Can such cancellation
also occur in higher order SWM? Now we can say yes. Using the
Feynman rules, the amplitude of $e^+(k_2)e^-(k_1)\to
\mu^+(p_1)\mu^-(p_2)$ can be written as

\begin{equation}
\begin{split}
M &=M_{\gamma}+M_{Z} \\
  &=\frac{ie^2}{s}\bar{v}(k_2)\gamma_{\mu}u(k_1)\bar{u}(p_2)\gamma^{\mu}v(p_1)e^{\frac{i}{2}(k_2\theta k_1+p_2\theta p_1)} \\
  & + \frac{ie^2}{\sin^2(2\theta_W)s}\bar{v}(k_2)\gamma_{\mu}(C_V
  - C_A\gamma_5)u(k_1)\bar{u}(p_2)\gamma^\mu(C_V -C_A\gamma_5)v(p_1)e^{\frac{i}{2}(k_2\theta k_1 + p_2\theta p_1)} \\
  &=M_{SM}e^{\frac{i}{2}(k_2\theta k_1+p_2\theta p_1)}
\end{split}
\end{equation}
where $M_{SM}$ is the amplitude in SM. Since the contributions
from SWM alone vanish due to the on-shell condition, the NC
correction merely appear as phase factors from the Moyal-Weyl
product, leading to no net noncommutative effect.

 \begin{figure}
 \includegraphics[scale = 0.5]{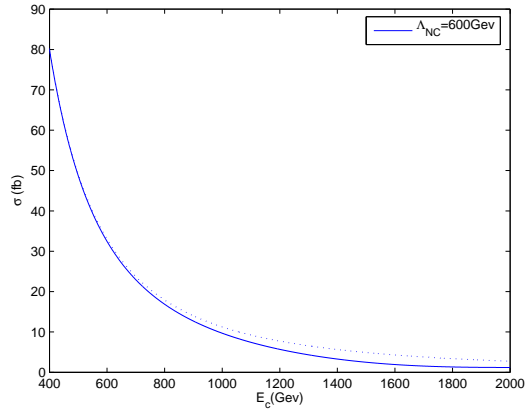}%
 \caption{The total cross section for $e^+e^-\to Z H$ as a function of $E_c$
 in ordinary SM (dotted line) and mNCSN with $\Lambda_{NC}$=600 \gev (solid line),
  $m_H$=135 \gev.}{\label{ttp1}}
 \end{figure}

\begin{figure}
\includegraphics[scale=.6]{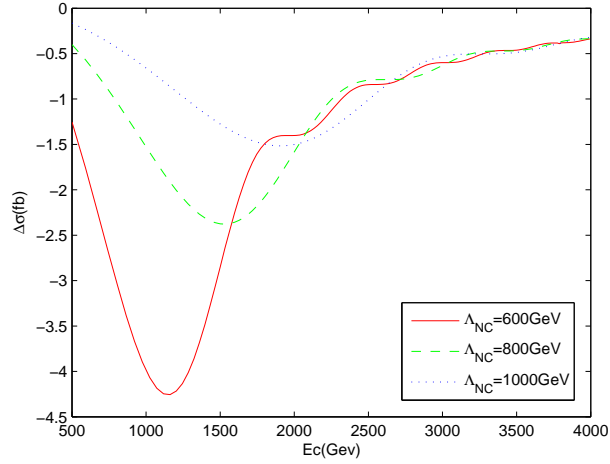}
\caption{The NC correction $(\Delta\sigma)_{NC}$ as a function of
  $E_c$ for $e^+e^-\to Z H$ with $m_H$=135 \gev, $\Lambda_{NC}$ = 600 \gev,
  800 \gev  and 1000 \gev, respectively.} \label{ttp2}
\end{figure}

\begin{figure}
 \includegraphics[scale = 0.56]{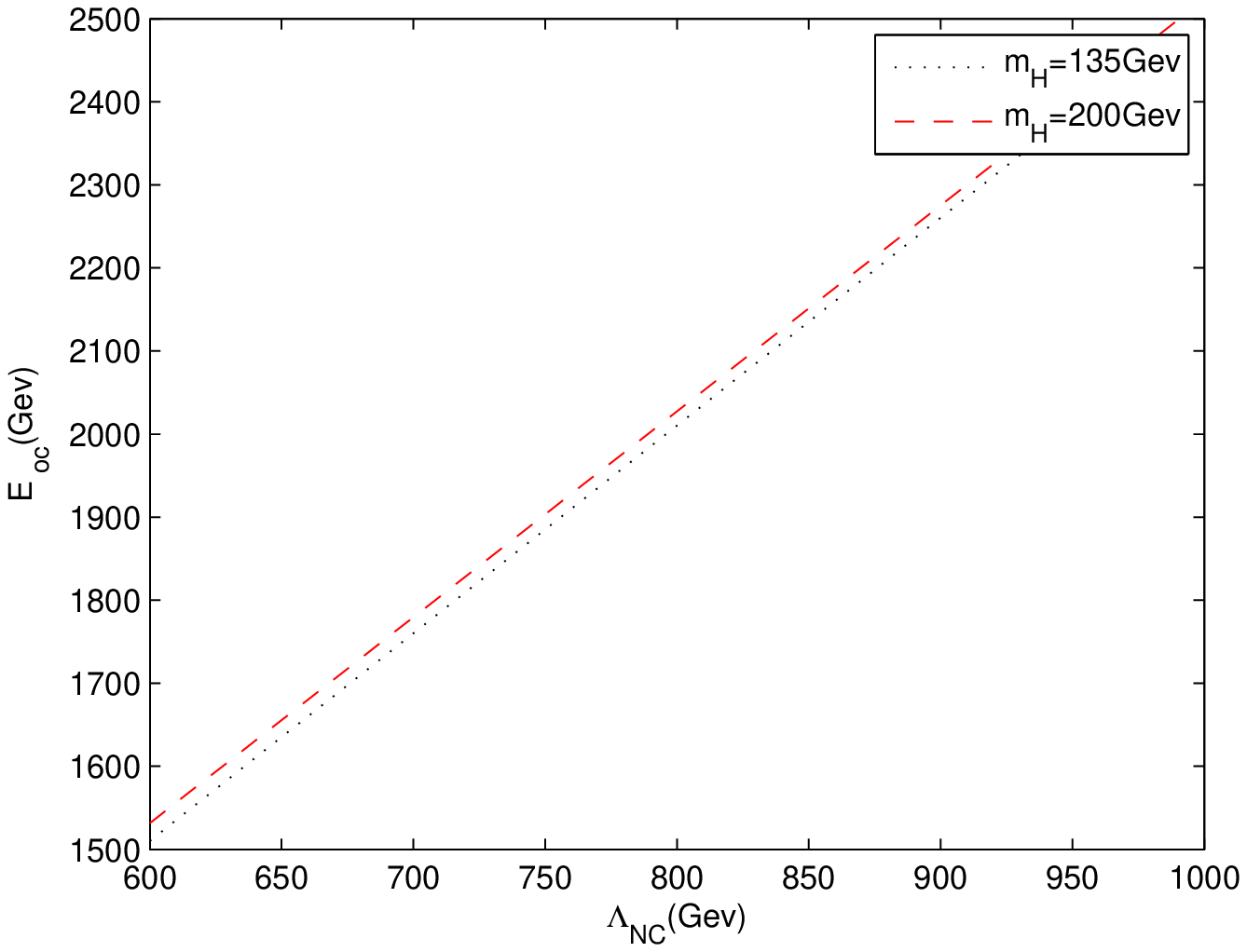}%
 \caption{The optimal collision energy $E_{oc}$ as a function of NC scale energy
  $\Lambda_{NC}$ for $e^+e^-\to Z H$ for $m_H$=135 \gev and 200 \gev, respectively.\label{ttp3}}
 \end{figure}

\begin{figure}
 \includegraphics[scale = 0.56]{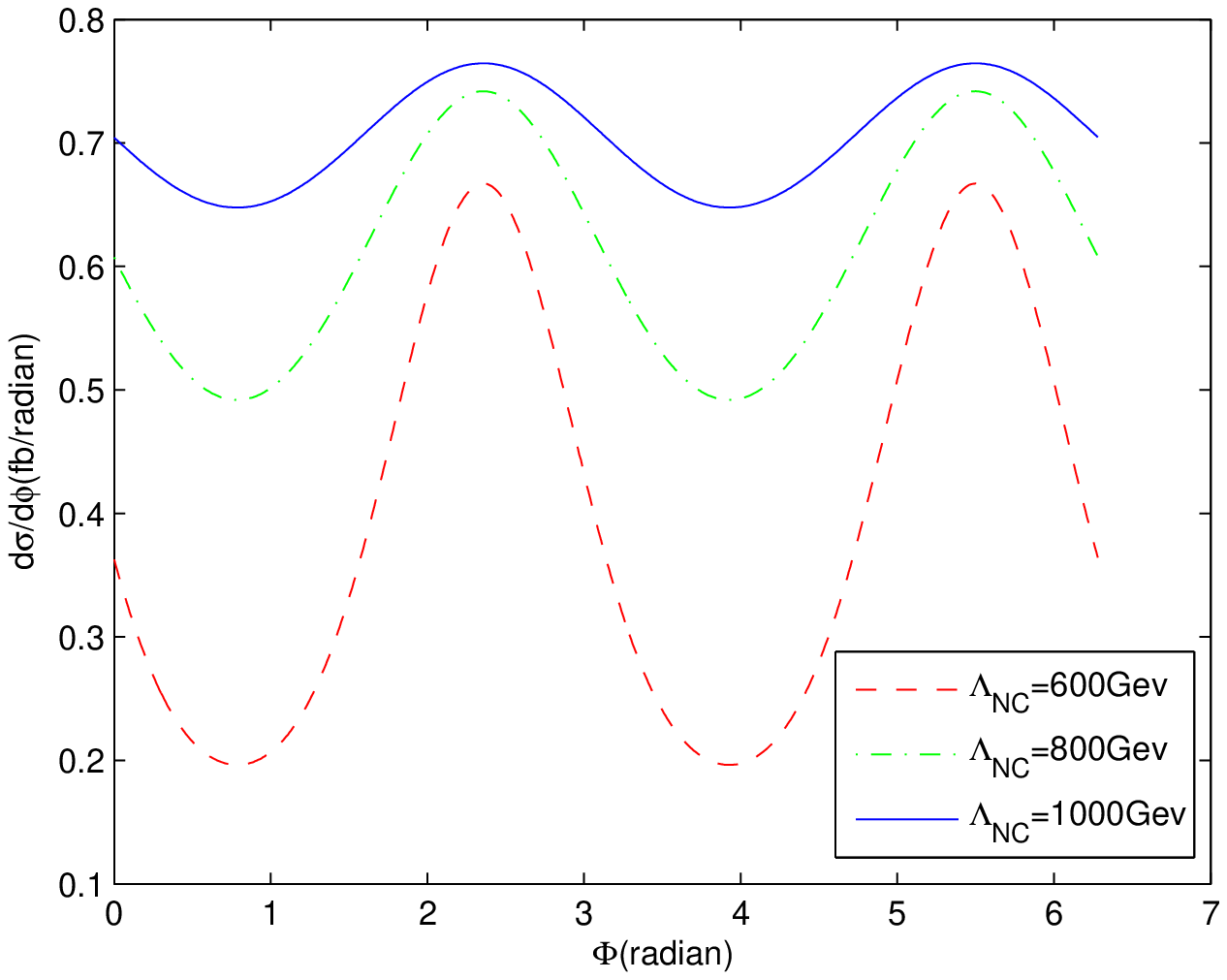}%
 \caption{$\frac{d\sigma}{d\Phi}$ as a function of $\Phi$ for $e^+e^-\to Z H$
 for $m_H$=135 \gev, $\Lambda_{NC}$=600 \gev, 800 \gev and 1000 \gev.}{\label{ttp3first}}
 \end{figure}

 \begin{figure}
 \includegraphics[scale = 0.53]{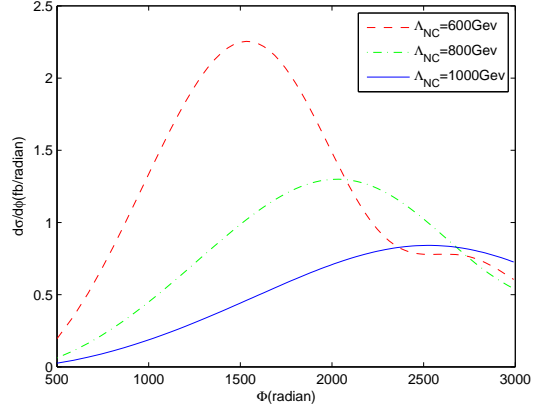}%
 \caption{The total cross section for $e^+e^-\to H H$ as a function of
 $E_c$,
for $m_H$=135 \gev, $\Lambda_{NC}$=600, 800 and 1000
\gev.}{\label{ttp04}}
 \end{figure}

 \begin{figure}
 \includegraphics[scale = 0.53]{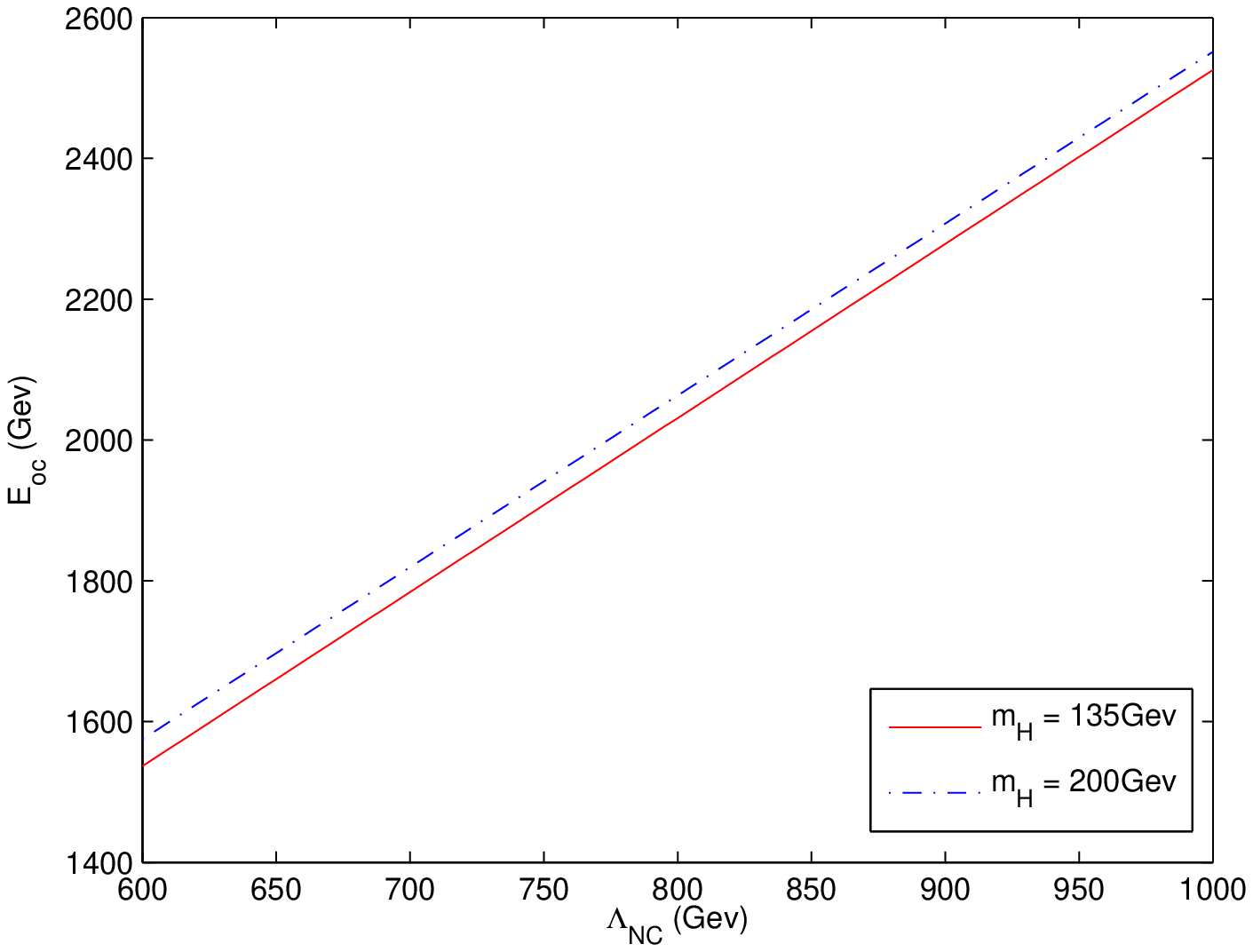}%
 \caption{The optimal collision energy $E_{oc}$ as a function of NC scale energy
  $\Lambda_{NC}$ for $e^+e^-\to H H$ for $m_H$=135 \gev and 200 \gev.\label{ttp23}}
 \end{figure}

\begin{figure}
 \includegraphics[scale = 0.53]{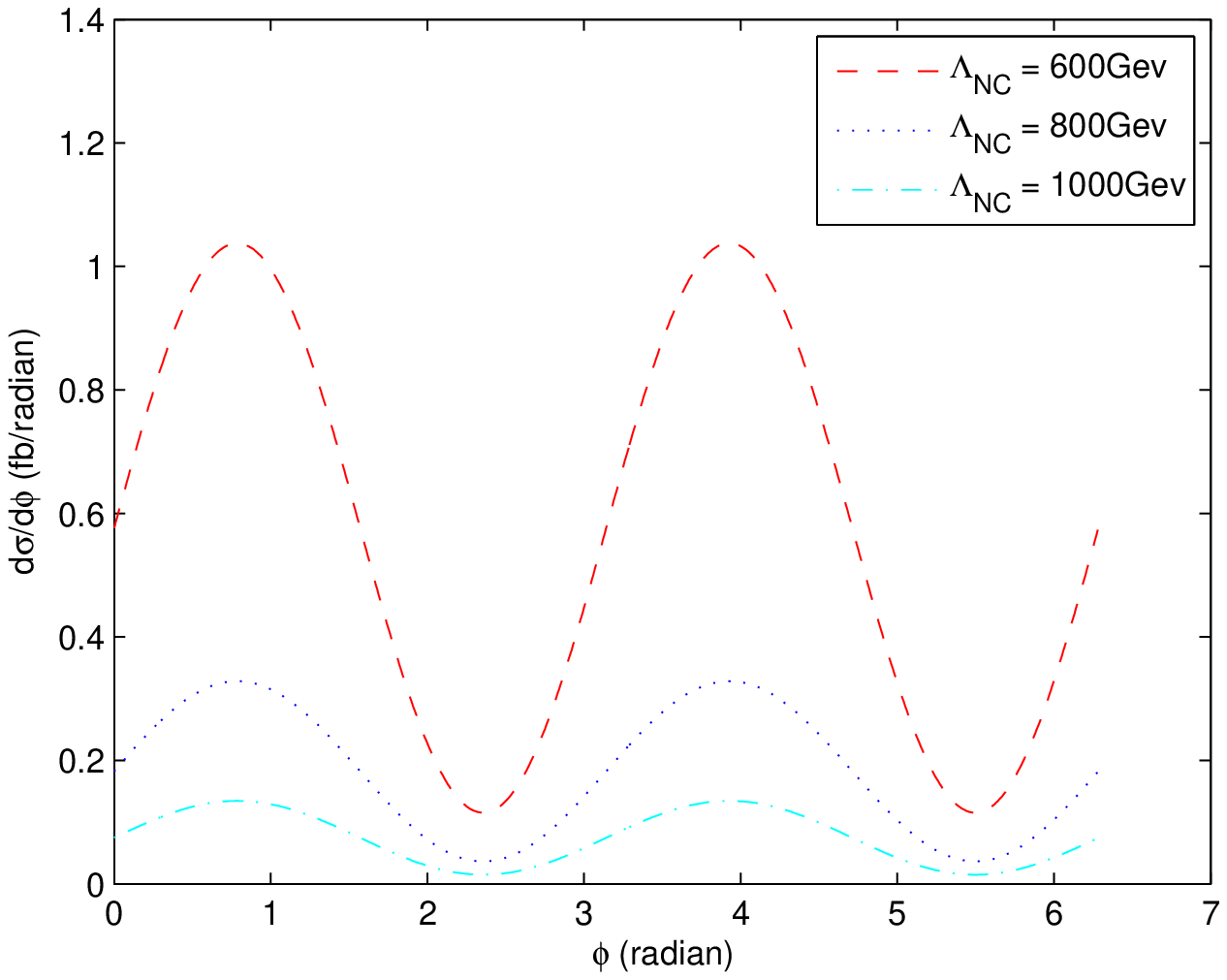}%
 \caption{The $\frac{d\sigma}{d\Phi}$ as a function of $\Phi$ for $e^+e^-\to H H$
  for $m_H$=135 \gev, $\Lambda_{NC}$=600 \gev, 800 \gev and 1000 \gev.}{\label{ttp05}}
 \end{figure}

  \begin{figure}
 \includegraphics[scale = 0.53]{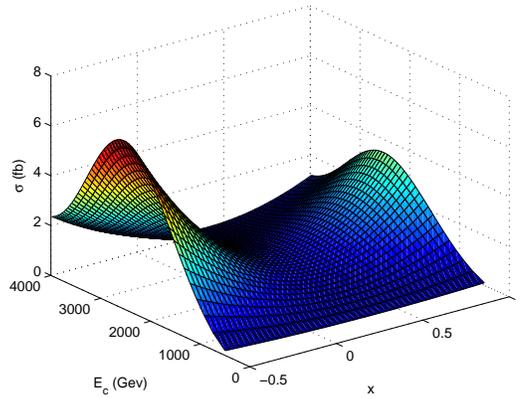}%
 \caption{The total cross section for $e^+ e^- \to H H$ as a function of $E_c$
 and x for $m_H$=135 \gev, $\Lambda_{NC}=1000 $ \gev.}{\label{ttp21}}
 \end{figure}

\section{NC CROSS SECTION AND NUMERICAL ANALYSIS}

The differential cross section for two body process is given by

\begin{equation}
\frac{d\sigma}{d\cos\theta
d\phi}=\frac{1}{64\pi^2s}\overline{|M|}^2
\end{equation}

where $\theta$ and $\phi$ are polar  and azimuthal angles,
respectively.
 Then the NC correction is

\begin{equation}
(\Delta\sigma)_{NC}=\sigma - \sigma_{0}
\end{equation}

where $\sigma_{0}$ is the total scattering cross section in
ordinary space-time. We are also interested in the relative
correction:
\begin{equation}
\delta_r = \frac{(\Delta\sigma)_{NC}}{\sigma_{0}}
\end{equation}

In the following analysis, we decompose $c_{\mu \nu}$ into
electric-like parts $\vec{\theta}_E = (\theta_{01},
\theta_{02},\theta_{03})$ and magnetic-like parts $\vec{\theta}_B
= (\theta_{23}, \theta_{31}, \theta_{12})$, where the vectors
$\vec{\theta}_E$ and $\vec{\theta}_B$ are given in Refs.
\cite{Das08,Ab10}, i.e.,
$\vec{\theta}_E=\frac{1}{\sqrt{3}}(\vec{i}+\vec{j}+\vec{k})$,
$\vec{\theta}_B=\frac{1}{\sqrt{3}}(\vec{i}+\vec{j}+\vec{k})$.

\subsection{Cross section and angular distribution of $e^+e^-\to Z H$ in NCSM}

In Fig. \ref{ttp1}, we show the ordinary total cross section
$\sigma_{0}$ and the NC corrected cross section $\sigma $ as
function of
 the collision energy $E_c(=\sqrt{s})$ for $m_H=135$ \gev and NC scale $\Lambda_{NC}=600$ \gev.
We can see from the figure that the NC effect significantly
suppresses the ordinary total cross section when $E_c$ is high
enough. In Table I, we present the relative correction for
$E_c=1000$ \gev and $1500$ \gev  for different parameters. The
$(\Delta\sigma)_{NC}$ as a function of the collision energy is
presented in Fig. \ref{ttp2}. The curve shows a negative kurtosis
distribution that has a maximum correction if the NC scale is
fixed. For $\Lambda_{NC}=600$ \gev, $800$ \gev, and $1000$ \gev,
the $(\Delta\sigma)_{NC}$ reaches its largest correction when the
collision energy is at 1500 \gev, 2000 \gev, and 2500 \gev,
respectively. It is useful to obtain a relation between the NC
scale energy $\Lambda_{NC}$ and the optimal collision energy
$E_{oc}$, as shown in Fig. \ref{ttp3}. We have
\begin{eqnarray}
E_{oc}&=&2.4986\Lambda_{NC}+7.9642 \quad\quad(m_H = 135 \gev)\\
E_{oc}&=&2.4789\Lambda_{NC}+44.2820\quad\quad(m_H = 200 \gev)
\end{eqnarray}

\begin{table}
  \caption{\label{}The relative correction for the process $e^+ e^-\to Z H$
   with collision energy $E_c$=1000 \gev and 1500 \gev; $m_H$=135 \gev and 200 \gev;
    $\Lambda_{NC}$ = 600 \gev, 800 \gev, and 1000 \gev, respectively.   }
  \begin{ruledtabular}
    \begin{tabular}{c|c|c|c}
     $E_C$(\gev)  &  $\Lambda_{NC}$(\gev)  &  $\delta_r(m_H=135$\gev)  &  $\delta_r(m_H=200$\gev)\\
     \hline
     1000        &  600                  &   0.1389                 &  0.1332              \\
                 &  800                  &   0.0468                 &  0.0447              \\
                 &  1000                 &   0.0195                 &  0.0186              \\
     1500        &  600                  &   0.4913                 &  0.4863              \\
                 &  800                  &   0.2161                 &  0.2125              \\
                 &  1000                 &   0.0968                 &  0.0950              \\
    \end{tabular}
  \end{ruledtabular}
\end{table}

When the Higgs boson mass is accurately measured in the LHC or
other devices, the relations given here can provide an effective
method for indirectly estimating the NC scale value since it is
much easier to determine the peak point of a curve than its
inflexion point.
 Similar relations were obtained in the context of NC QED \cite{Sheng07}. In the mNCSM scenario,
one cannot get such a linear relation by simply expanding the
Lagrangian to the $\theta$ order. As  shown in many papers, in
this case the NC scattering cross section changes monotonously
when the collision energy is gradually increased.

We show the azimuthal angular distribution $\frac{d\sigma}{d\phi}$
in Fig. \ref{ttp3first}.
 Here the collision energy $E_c$ is $1.5$ \tev. One can see from
the figure that $\frac{d\sigma}{d\phi}$ is anisotropic. This is
due to an inherent characteristic of NC space-time. The curves
reach their maxima at $\phi=2.37$ rad and $\phi=5.51$ rad. The two
minima  are located at $\phi=0.80$ rad and $\phi=3.95$ rad. This
unique feature can help us in identifying the NC effect from the
other effects.

\subsection{Cross section and angular distribution of $e^+e^-\to H H$ in NCSM}

The neutral Higgs pair production $e^+e^-\to H H$ is forbidden at
the tree level in ordinary standard model.
 Thus the correction in the cross section is just the cross section itself.
The reason why we are particularly interested in the process is
that any signal of a SM forbidden process will imply new physics.

For simplicity, we first set $x=\frac{1}{2}$ in Eqs.\eqref{Stacy}
and \eqref{Marshall}, which is corresponding to the case as that
in Ref.\cite{Calmet07}, \ie, the process is only Z mediated. The
total cross section $\sigma$ as a function of collision energy
$E_{c}(=\sqrt{s})$ is shown in Fig. \ref{ttp04}. Here we set $m_H
= 135$ \gev  with NC scale $\Lambda_{NC} = 600$ \gev, $800$ \gev
and $1000$ \gev. As expected, a maximum cross section appears. The
relative optional collision energy is located at about $1500$
\gev, $2000$ \gev and $2500$ \gev for the cross sections  $2.25$
fb, $1.30$ fb, and $0.84$ fb, respectively.

The relation between $E_{oc}$ and $\Lambda_{NC}$ is given by
\begin{eqnarray}
E_{oc}&=&2.4728\Lambda_{NC}+52.9257 \quad\quad(m_H = 135Gev)\\
E_{oc}&=&2.4414\Lambda_{NC}+110.2084\quad\quad(m_H = 200Gev)
\end{eqnarray}
as shown in Fig. \ref{ttp23}

The azimuthal angular distribution $\frac{d\sigma}{d\phi}$ is
given in Fig. \ref{ttp05} for $m_H=135$ \gev. The curves are for
$\Lambda_{NC} = 600$ \gev, $800$ \gev, and $1000$ \gev,
respectively. The maxima (minima) are at $\phi =$ 2.36 rad, 5.50
rad (0.79 rad, 3.93 rad), respectively.

Now we consider the impact of the U(1) gauge ambiguity discussed
in Sec. 2, which does not contribute to $e^+e^- \to Z H$. In this
case, the contribution from photon-Higgs-Higgs diagram must be
considered. Using Eqs. \eqref{tjack}, \eqref{john} and
\eqref{sayid} we obtain and show in Fig. \ref{ttp21} the total
cross section as a function of $E_c$ and $x$ for $m_H = 135$ \gev
and $\Lambda_{NC} = 1000$ \gev. Here we assume that $x$ varies
between $-0.5$ and $1$. One can see that the cross section shows a
parabolic dependence on $x$ when the collision energy is  fixed.
The saddle point in Fig. \ref{ttp21} is at $x=0.4$. When $x$ is
located at [0.5, 1] or [-0.5, 0.3], the total cross section is
greatly enhanced. However, if $x$ is in [0.3, 0.5], the cross
section will be slightly suppressed.

\section{CONCLUSION AND DISCUSSION}

In this paper we have explored the NC effect in the Higgs boson
production process $e^+e^-\to Z H$ and SM forbidden process
 $e^+e^-\to H H$. Several new results are obtained. First,
the n-th order Seiberg-Witten map for complex scalar fields is
given. Despite the lengthy expression, for the processes discussed
we can still obtain enough information to get the complete Feynman
rules. Second, it is found that the NC effect can significantly
reduce the cross section of the process $e^+e^-\to ZH$ when the
collision energy exceeds 1 \tev. For $e^+e^-\to H H$, we obtained
the total cross section and angular distribution using the
simplest representation of SWM given by Ref. \cite{Calmet07}.
Moreover, we can also include more complicated representation, as
well as photon-Higgs-Higgs interaction which does not arise in
Ref. \cite{Calmet07}. It is shown that although the process
$e^+e^-\to ZH$ is independent from this changing, the total cross
section of $e^+e^-\to H H$ cross section can be enhanced. This
increases our confidence for detecting the NC signal associated
with the of Higgs boson, in the future International Linear
Collider. For each process
 we can find an optimal collision energy as a function of the NC scale
 $\Lambda_{NC}$ in order to get the largest NC correction, which can help us
 to determine $\Lambda_{NC}$ effectively. Finally we
briefly comment on the process $e^+e^-\to \mu^+ \mu^-$ studied in
Ref. \cite{Ab10}. Using the n-th order Seiberg-Witten map, we show
that the NC scattering amplitude differs from the ordinary one by
only a phase factor, without NC effect.

The SWMs given in Sec. II are not the general. One can add an
homogeneous solution of the Seiberg-Witten equation to obtain
another solution. As is well known, the degrees of freedom play an
essential role in the renormalization of NCQFT
\cite{Bichl01,Buric02}. For the process $e^+e^- \to H H$ and
$e^+e^- \to\mu^+ \mu^-$, all these ambiguity vanish because of the
on-shell condition, thus the physical results are freedom
independent. For the process $e^+e^-\to Z H$, the contributions
from the homogeneous solutions containing two gauge fields cancel
or vanish when the on-shell condition is applied. Thus the
contribution from these degrees of freedom is limited to that
containing one gauge field. Until now the phenomenological
modification of these homogeneous solutions have not been
considered, except for the pure gauge sector \cite{Buric07}. This
is because we still do not have enough information on the
renormalizability of NCQFT. We expect that further progress on the
renormalizability of the noncommutative Higgs sector can finally
remove this ambiguity and provide a more solid foundation for the
phenomenological study, as has been done in the pure gauge sector.
It should be noted that a compromising and more practical method
is given in Ref. \cite{Tamarit09} where all possible deformed
terms were considered. In any case, here we have demonstrated the
rich phenomenological correlation between the Higgs physics and
noncommutative spacetime, and  other important production
processes such as $e^+ e^-\to \bar{\nu}_e \nu_e H$ are being
investigated.

\nn{\bf Acknowledgments:}

The authors would like to thank Prof. M.Y YU for his useful
discussions. This work is supported in part by the funds from NSFC
under Grant No.11075140 and the Fundamental Research Funds for the
Central University.

\appendix*
\section{THE FEYNMAN RULE FOR Z-H-H INTERACTION}

The Feynman rules for the Z($k$)-H($p_{1}$)-H($p_{2}$) vertex
given in Ref. \cite{Das11} is
\begin{equation}
\frac{gm_{H}^2(k\Theta)_\mu}{4\cos\theta_W} \label{eq1}
\end{equation}
Since the expression \eqref{eq1} is proportional to $m_{H}^2$, we
need only to investigate
\begin{equation}
\int
d^4x((\partial_\mu\hat{\Phi}^{\dagger})*(\partial^\mu\hat\Phi) -
\mu^2\hat\Phi*\hat\Phi - \lambda(\hat{\Phi}^\dagger*\hat{\Phi})^2)
\label{eq2}
\end{equation}
in the Higgs sector.

Following Ref. \cite{Calmet07}, we take the SWM representation
\begin{equation}
\hat{\Phi} =
\Phi-\frac{1}{2}\theta^{\alpha\beta}V_\alpha\partial_\beta\Phi -
\frac{i}{4}\theta^{\alpha\beta}V_{\beta}(V_\alpha\Phi - \Phi
V'_\alpha), \label{eq3}
\end{equation}
where
\begin{equation}
V_{\mu} = \frac{1}{2}g'B_{\mu} +
gW_\mu^a\frac{\sigma^a}{2}=\left(\!\!\!\begin{array}{cc}
eA_\mu+\frac{g}{2\cos\theta_{W}}(1-2\sin^2\theta_W)Z_\mu&\frac{g}{\sqrt{2}}W_\mu^+\\
\frac{g}{\sqrt{2}}W_\mu^-&-\frac{g}{2\cos\theta_W
}Z_\mu\end{array}\!\!\!\right).
\end{equation}
The last term of RH in Eq. \eqref{eq3} contains two gauge fields,
which is not related to the Z-H-H interaction. Excluding this
term, one has
\begin{equation}
\hat{\Phi} =
\Phi-\frac{1}{2}\theta^{\alpha\beta}V_\alpha\partial_\beta\Phi
\end{equation}
When
$\Phi\longrightarrow{SSB}\frac{1}{\sqrt{2}}\left(\!\!\!\begin{array}{c}0\\v+h
\end{array}\!\!\!\right)$, we have
\begin{equation}
\hat{\Phi}\to \frac{1}{\sqrt{2}}\left(\!\!\!\begin{array}{c}0\\ v
+ h \end{array}\!\!\!\right) -
\frac{1}{2}\theta^{\alpha\beta}V_\alpha\frac{1}{\sqrt{2}}\left(\!\!\!\begin{array}{c}
0 \\ \partial_\beta h \end{array}\!\!\!\right) =
\frac{1}{\sqrt{2}}\left(\!\!\!\begin{array}{c}
-\frac{g}{2\sqrt{2}}\theta^{\alpha\beta}W_\alpha^+\partial_\beta
h\\ v + \hat{h}  \end{array}\!\!\!\right) \label{eq4}
\end{equation}
where
\begin{equation}
\hat{h}= h + \frac{\theta^{\alpha\beta
}g}{4\cos\theta_W}Z_\alpha\partial_\beta h
\end{equation}

For simplicity we rewrite it as

\begin{equation}
\hat{h}= h + \theta f. \label{eq5}
\end{equation}

Inserting Eq. \eqref{eq4} into \eqref{eq2} and ignoring the
unrelated terms, we obtain
\begin{equation}
\int
dx^4(\frac{1}{2}(\partial_\mu\hat{h}^{\dagger})(\partial^\mu\hat{h})
-\frac{\lambda}{4}v^{2}(\hat{h}^{\dagger}\hat{h}^{\dagger}+2\hat{h}^{\dagger}
\hat{h}+\hat{h}\hat{h}))
\end{equation}

Using \eqref{eq5} and taking partial integration, the
corresponding NC correction up to the $\theta$ order is given by
\begin{equation}
-\theta\int dx^4 f (\partial_\mu\partial^{\mu}h + m_H^2h).
\end{equation}
Obviously, when the Higgs boson is on-shell, this term does not
contribute to the Z-H-H vertex, thus leaving no NC effect.

\end{document}